\documentclass[aps,prl,twocolumn,groupedaddress,showpacs]{revtex4}

\usepackage{amssymb}
\usepackage{graphicx}
\usepackage{amsmath}
\usepackage[colorlinks=true,linkcolor=RoyalBlue,citecolor=OliveGreen,linktoc=page]{hyperref}
\usepackage[usenames,dvipsnames]{color}

\begin{document}

\title{Numerical analysis of second law of thermodynamics and irreversibility in exemplary quantum systems}

\author{
	G.B.~Lesovik$^1$ and
	I.A.~Sadovskyy$^{2,3}$
}

\affiliation{
	$^1$L.D.~Landau Institute for Theoretical Physics RAS, 
	Akad. Semenova av., 1-A, 142432, Chernogolovka, Moscow Region, Russia
}

\affiliation{
	$^2$Rutgers University,
	136 Frelinghuysen Road, Piscataway, New Jersey 08854, USA
}

\affiliation{
	$^3$Materials Science Division, Argonne National Laboratory,
	9700 S. Cass Avenue, Argonne, Illinois 60637, USA
}

\date{\today}

\begin{abstract}
We test Boltzmann's H-theorem for several models of particle random walk. We study the influence of interaction between the particle and reservoir/detectors on entropy and find entropy increasing in time for some models and behaving non-monotonically for others. The key mechanism affecting the entropy growth is the quantum entanglement between the system and the reservoir. We discuss the details of the system-reservoir interaction, such as presence of the interference in the system and number of interactions with detector parts, and their impact on the monotonicity of entropy.
\end{abstract}


\maketitle

{\it Introduction.} The second law of thermodynamics is the fundamental law of nature~\cite{Callen:1985}. One of its well-known formulations assumes the existence of state function (entropy), which does not decrease for isolated thermodynamic systems. This formulation implies the irreversibility in thermodynamic systems. At low temperatures quantum nature manifests itself~\cite{Spicka:2005,Capek:2005} and the second law of the thermodynamics may not necessarily work in its classical formulation. For example, let us consider a standalone quantum system and define the entropy $S$ according to von Neumann $S = -{\rm Tr}\{{\hat\rho}\log{\hat\rho}\}$, where $\hat\rho$ is the density matrix of the system. In this case the entropy of this system neither increases nor decreases, but remains constant~\cite{Nielsen:2004}.
Then the question arises: what kind of non-isolated quantum systems reveals the entropy growth and what is the mechanism of this behavior from the quantum mechanical point of view?

This question is particularly relevant for the quantum thermodynamics in small systems. Some of them were considered in Refs.~\cite{Kim:2007,Kim:2006,Ford:2006,Ford:2005,Buttiker:2005,Hanggi:2006}.
Recently several articles have been published~\cite{Gemmer:2001,Gemmer:2004,Hartmann:2005,Winter:2006}, where entropy growth was discussed from the quantum mechanical point of view.
In these articles the key qualitative point is the entanglement between the system and the reservoir. Moreover, there is a quantitative statement, that small subsystem interacting with the large quantum system obeys the entropy growth in the majority number of cases, made in Refs.~\cite{Gemmer:2001,Gemmer:2004}. Later the same authors have found that the small subsystem does not necessarily relax to the equilibrium state~\cite{Hartmann:2005}, hence the deviation from the standard thermodynamic behavior is implied. Thus, it remains unclear what types of reservoirs provide the entropy growth.

However, there is no doubt that during the interaction the energy exchange (heat exchange in terms of thermodynamics) should be infinitesimally small. Otherwise the entropy can either increase or decrease depending on the heat flux.

\begin{figure}[b]
   \includegraphics[width=7.8cm]{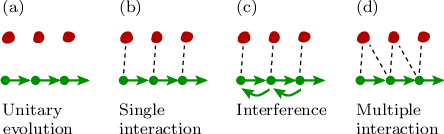}
   \caption{
Quantum mechanical system (green) connected to the reservoir (red). System evolution is schematically shown by green arrows. For clarity the reservoir is divided into several independent parts. The interaction is marked by a dashed line.
(a)~Non-interacting system evolves unitarily and its entropy does not change.
(b)~``Single shot'' interaction: the system interacts with a new part of the reservoir at every step.
(c)~System can interfere with itself.
(d)~System can interact with some parts of the reservoir several times.
   }
   \label{fig:system_reservoir}
\end{figure}

In this article we discuss a random walk of a quantum particle, which in some cases can be described by the kinetic (master) equation, but in some other cases can be not. We study several model systems interacting with reservoirs (see Fig.~\ref{fig:system_reservoir}) and determine the condition for the entropy growth. We focus on the two important features possibly causing the deviation from the kinetic equation.

The first feature is the quantum interference in the system. Usually the entropy of non-coherent systems tends to grow with time. However, coherent systems with interference [see Fig.~\ref{fig:system_reservoir}(c)] can not be described by the master equation for probabilities and the H-theorem~\cite{Tolman:1979} is not directly applicable for such systems. Therefore, in principle the quantum interference might lead to the entropy decrease during some special interactions with the reservoir.

Another source of the entropy decrease is the multiple interactions with the reservoir, see Fig.~\ref{fig:system_reservoir}(d).
This situation is quite special: usually only single interaction with the reservoir occurs, e.g., photons (phonons) after interacting with a system disappear in the reservoir and/or lose coherence. In case of multiple interactions the ``working'' parts of the reservoir should be in the vicinity of the system all the time. To comply with these requirements we consider one-dimensional system equipped by an array of spins acting as a reservoir in this article.

Technically, we use the scattering matrix approach for a random walk of a quantum particle~\cite{Feldman:2004}. In contrast to the classical random walk, the master equation for probabilities can not be used in this case. Note that random walk of a quantum particle described here differs from the so called quantum random walk~\cite{Aharonov:1993,Childs:2002}, where tunneling probabilities depend on the state of extra degree of freedom; in our case dependence is due to the phase difference accumulating over different Feynman paths (particle trajectories) only.
First we describe the one-dimensional system with scatterers and spins located along this system. The electron can be transmitted or reflected on scatterers keeping phase coherence and rotates each spin when passing it. We observe the evolution of the system state in discrete time and found the entropy sometimes decreases during some specific interactions with the reservoir.
The second system we consider is a binomial tree system with no back reflection from beam splitters on the nodes. An electron enters to the root, is split in half for each node, and acts on the new spin each next step. There is no interference in this system and it interacts with a given part of the reservoir (given spin) just once. Expectedly, the entropy of this system monotonically increases in time. In the case of complete phase breaking after each step (e.g., each spin rotates by $\pi$) the system can be described by a kinetic equation and the H-theorem can be proven explicitly.
The third system demonstrates an intermediate case. It is a tree-like system with interference and single interaction with reservoir parts.

\begin{figure}[tb]
	\includegraphics[width=3.8cm]{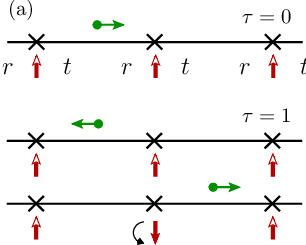} $\;$
	\includegraphics[width=4.2cm]{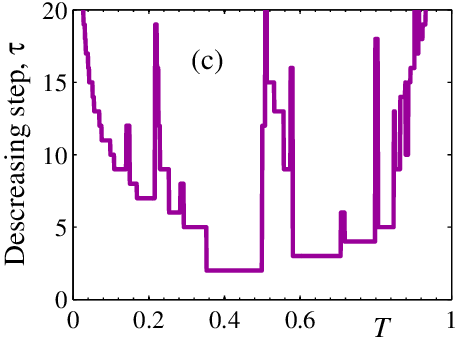} \newline
	\includegraphics[width=8.3cm]{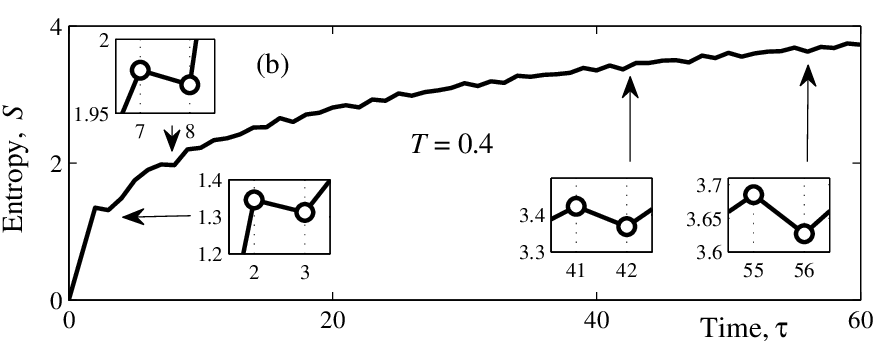} \newline
	\includegraphics[width=8.3cm]{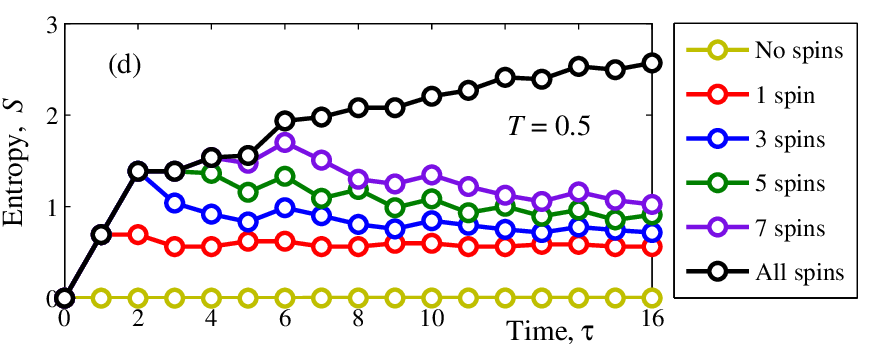}
	\caption{
(a)~One-dimensional system with scatterers (crosses) with transmission and reflection amplitudes $t$ and $r$. Electron (green dot) starts to move right at position $i=0$ between scatterers and $\tau=0$. Each spin (red arrow) rotates by $\pi$ when electron moves near it. Two possibilities at $\tau=1$ are shown: transmitted electron moved near spin once and have rotated it, while reflected electron formally moved near spin even (0 or 2) times and have left it in the same position.
(b)~Entropy of one-dimensional model as a function of time $\tau$ ($t = \sqrt{T}$, $r = i\sqrt{1-T}$, and $T=0.4$) monotonically increases ``in average''. Small insets show the non-monotonic behavior between some steps, e.g., $\tau = 2$-3, 7-8, 41-42, 55-56, etc.
(c)~First step $\tau$ followed by entropy drop plotted as a function of transparency $T=|t|^2$.
(d)~Entropy as a function of $\tau$ for several spins symmetrically located at and around first scatterer [center cross in Fig.~(a)].
	}
	\label{fig:1d}
\end{figure}

{\it Quantum one-dimensional walk.} Let us consider an ideal one-dimensional wire with equivalent scatterers located at equal distances from each other, see Fig.~\ref{fig:1d}(a). Each scatterer has a transmission amplitude~$t$ and reflection amplitude~$r$. We assume non-smearing wave package (electron) moving with fixed speed and introduce discrete time $\tau$ enumerating the moments when the wave package is located exactly between two neighbor scatterers. 

The reservoir is presented by spins located at scatterers. Two possible processes at the first scatterer are shown in Fig.~\ref{fig:1d}(a). Initially ($\tau=0$) the electron is located exactly between two scatterers ($i=0$) and moves to the right. At the first step ($\tau=1$) the reflected electron is located at the same position and moving left (with probability $R=1-T=|r|^2$), whereas the transmitted electron is located right to the initial position and moving right (with probability $T=|t|^2$). 
The system-reservoir interaction is given by spin rotating by some certain angle when the electron passes through the scatterer and remaining in the same state if the electron is reflected from the scatterer. For simplicity this angle is chosen to be $\pi$. In this case the array of spins ``traces'' the electron position, but does not remember the history of its movements.
Performing this procedure at the next steps we obtain a set of probabilities $\{P_i^\rightarrow(\tau)\}$ ($\{P_i^\leftarrow(\tau)\}$) to observe the electron at position $i$ moving right (left). The probability to find the particle at position $i$ is $P_i(\tau) = P_i^\rightarrow(\tau) + P_i^\leftarrow(\tau)$. Note that due to the chosen initial conditions, $P_i^\leftarrow(\tau)=0$ if $i+\tau$ is even, $P_i^\rightarrow(\tau)=0$ if $i+\tau$ is odd. Therefore, $P_i(\tau)$ is defined only by the electron moving right or left and never by its superposition.

In this simple model the density matrix ${\hat\rho}_{\rm e} = {\rm Tr}_{\rm s}\{{\hat\rho}\}$ of the electron subsystem (the whole density matrix ${\hat\rho}$ traced over spin degrees of freedom) is already diagonal, ${\hat\rho}_{\rm e} = {\rm diag}\{P_0, P_1, \ldots\}$. This allows us to immediately find the von Neumann entropy $S = -\sum_i P_i \log P_i$. The entropy as a function of discrete time (steps)~$\tau$ is shown in Fig.~\ref{fig:1d}(b) for $T=0.4$. In average the entropy logarithmically increases in time, $S(\tau) \propto \log\tau$ (system parameters, such as transmission amplitude $t$, can slightly change the dependence). The system is strongly mixed and the entropy is close to its possible maximum.
On the other hand at some steps the entropy decreases, which demonstrates impossibility to extend H-theorem for such models. Figure~\ref{fig:1d}(c) shows the first time step $\tau$ followed by entropy decrease as a function of transmission probability $T$. For example, the first entropy drop occurs between 2\textsuperscript{nd} and 3\textsuperscript{rd} steps at $T \in [0.35 \ldots 0.50]$, between 3\textsuperscript{rd} and 4\textsuperscript{th} steps at $T \in [0.58 \ldots 0.71]$, and between 4\textsuperscript{th} and 5\textsuperscript{th} steps at $T \in [0.72 \ldots 0.80]$.

Here the electron and spin subsystems are treated as a system and reservoir respectively, see Fig.~\ref{fig:system_reservoir}. Note, since the whole system is pure, the entropies of the electron and spin subsystems are equal and therefore it is possible to present the electron subsystem as a reservoir.

Another important example we consider is one-dimensional motion with several spins. To be specific, we take odd number of spins symmetrically located at and around first scatterer. The entropy as a function of time $\tau$ is presented in Fig.~\ref{fig:1d}(d) for 1, 3, 5, and 7 spins. In the absence of spins the evolution is unitary and entropy remains zero. In the presence of all spins the entropy increases logarithmically in average as before (black line). In case of several spins the entropy behaves exactly as in the case of all spins at small $\tau$ and saturates at constant value $S_\infty = P_{\rm L}\log P_{\rm L} + P_{\rm R}\log P_{\rm R}$ at large $\tau$, where $P_{\rm L}$ and $P_{\rm R}$ are probabilities to observe particle at the left and at the right from the set of several spins ($P_{\rm L} + P_{\rm R} = 1$).

\begin{figure}[tb]
	\includegraphics[width=2.9cm]{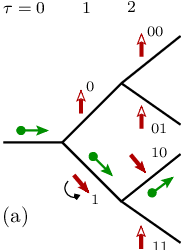} $\;$ %width=3.1cm
	\includegraphics[width=5.0cm]{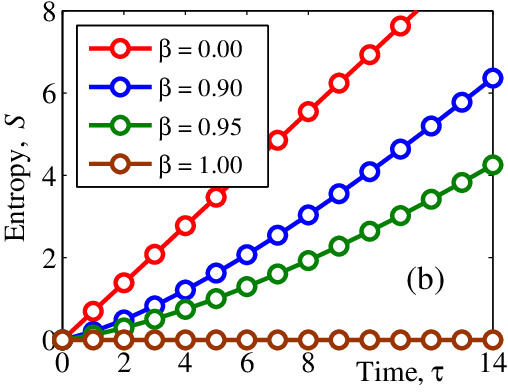}
	\caption{
(a)~Tree-like model setup without interference. The beam splitters without back reflection are located at all nodes. Particle can penetrate to lower or upper arm with 50/50 probability and can not be reflected back. Spins are located on each edge and rotate by some certain angle when electron moves through this edge. State of spins' subsystem has one-to-one correspondence to the electron path, e.g., spin at position with index $\chi = 10$ corresponds to the electron path shown in green arrows.
(b)~Entropy as a function of time $\tau$ for tree model without interference and $Z = 2$ monotonically increases in time.
}
	\label{fig:tree}
\end{figure}

{\it Binomial tree without interference.} Let us consider a tree with scatterers at the nodes and ideally conducting edges as shown in Fig.~\ref{fig:tree}(a). Each scatterer splits the incoming wave packet to the two outgoing wave packets without backscattering. It can be described by $3\times 3$ scattering matrix with ``transmission''~$t = \sqrt{T}$ (incoming to the first outgoing) and ``reflection''~$r = i\sqrt{1-T}$ (incoming to the second outgoing) amplitudes.

We again assume no wave packet spreading and introduce discrete time $\tau$. At $\tau = 0$ the wave packet enters to the root edge, at $\tau = 1$ the wave packet is split once and two resulting wave packets are located at the next two edges, at $\tau = 2$ the wave packets are located at four edges, and so on.

In this model the reservoir/detectors are presented by spins at the edges. The spin rotates by some certain angle~$\varphi$ while electron is moving through this edge. We enumerate the spins in the following way: two spins $\chi_0$ and $\chi_1$ are located after the first splitter; four spins $\chi_{00}$, $\chi_{01}$, $\chi_{10}$, and $\chi_{11}$ after the next two splitters, etc. At the beginning, all spins are in some initial state $\chi_{\zeta,\rm i}$, which becomes $\chi_{\zeta,\rm f}$ after rotating by the angle~$\varphi$, where~$\zeta$ is the $\tau$-digit spin index ($\zeta = 0, 1, 00, 01, 10, 11, \ldots$). The spin index contain all information about electron path, e.g., $\zeta = 010$ corresponds to the electron path $0 \to 1 \to 0$.
We introduce a phenomenological constant $\langle \chi_{\zeta,\rm f} \chi_{\zeta,\rm i}^\dag \rangle = \alpha$, the same for all spins. It may also serve, e.g., for approximate description of decoherence due to the breaking radiation. Naturally, $\langle \chi_{\zeta,\rm i} \chi_{\zeta',\rm i}^\dag \rangle = \langle \chi_{\zeta,\rm f} \chi_{\zeta',\rm f}^\dag \rangle = \delta_{\zeta\zeta'}$ and $\langle \chi_{\zeta,\rm i} \chi_{\zeta',\rm f}^\dag \rangle = \alpha\delta_{\zeta\zeta'}$. The density matrix can be expressed in terms of $\alpha\alpha^*$, thus we will use $\beta \equiv \alpha\alpha^*$ as a braking radiation parameter below.

Wave function of the electron and spin subsystems defines state of the system. Using this wave function we calculate the density matrix then take the trace over the spin degrees of freedom, which results in $\tau\times\tau$ electron density matrix at $\tau$ step.
The entropy for double output splitters ($Z = 2$) and $T = 0.5$ as a function of~$\tau$ is shown in Fig.~\ref{fig:tree}(b). It monotonically increases in time, thus we would like to claim that the H-theorem can be broadened to quantum systems without interference and with single system-reservoir interaction, though the density matrix of these systems is non-diagonal in contrast to the kinetic equation case.
We have tested it for different $\beta$'s, different number of splitter outputs $Z > 2$, and different transmission amplitudes of each beam splitter (keeping its unitarity). Also we considered different models of tree without interference, e.g., model where only one channel splits at every step. The qualitative result is the same: entropy monotonically increases with~$\tau$.

The entropy growth occurs due to the increasing entanglement with the spin subsystem acting as a reservoir. The details depend on how new degrees of freedom comes into play: roughly, quicker entanglement leads to faster entropy growth.
In the case of complete phase breaking after the system density matrix becomes diagonal. It diagonal elements obey the master equation and satisfy the H-theorem conditions. The keynote is the splitting of probability $P$ at step $\tau$ into two parts with probabilities $P_1$ and $P_2 = P - P_1$ at step $\tau + 1$. The entropy of the corresponding part is non-decreasing $s(\tau+1) = -P_1\log P_1 - P_2 \log P_2 \geqslant -P_1\log P - P_2 \log P = -P \log P = s(\tau)$, thus the total entropy is also non-decreasing.

\begin{figure}[tb]
	\includegraphics[width=3.9cm]{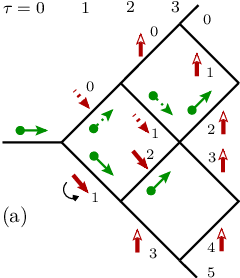} \vspace{1mm} $\;$
	\includegraphics[width=4.45cm]{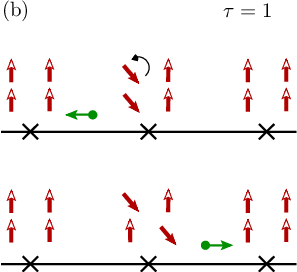}
	\includegraphics[width=4.15cm]{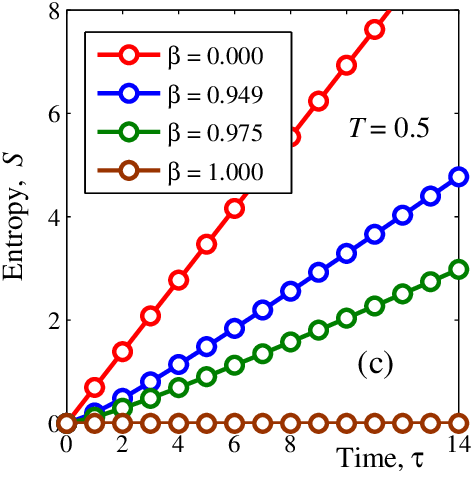} $\;$
	\includegraphics[width=4.15cm]{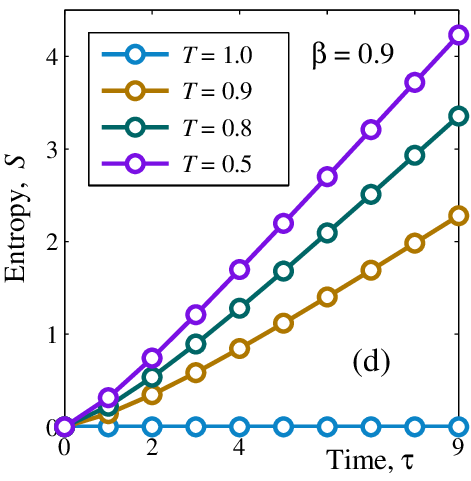}
	\caption{
(a)~Tree-like model setup with interference. Each no-back-refection beam splitter has two inputs and two outputs with amplitudes $t$ for transmitted and $r$ for deviated waves (unused inputs at the beginnings and edges are not shown). Spins located on each edge ``trace'' the electron path. Electron can interfere due to the loops between its paths.
(b)~Equivalent representation of the model in Fig.~(a). The one-dimensional motion with doubled spin set (before and after scatterer). Each spin rotates by some certain angle while electron moves near it. All these sets are ``remembered'' after each interaction and the new set of spins in initial positions is ``presetted'' for the each next interaction. For example, if the electron is reflected at the first scatterer, the spin in the first set rotates by some certain angle before scattering and the equivalent spin in the second set rotates by the same angle. If the electron is transmitted through the scatterer, then these positions of rotated spins in the first and the second sets are different.
(c)~Entropy as a function of time $\tau$ for tree model with interference, $Z = 2$, $T=0.5$, and different $\beta$.
(d)~Entropy on time $\tau$ for different $T = |t|^2$ and $\beta=0.9$.
}
	\label{fig:tree_interference}
\end{figure}

{\it Binomial tree with interference.} Now let us consider an intermediate case. The system with interference can interact with a given spin only once. We define a tree as shown in Fig.~\ref{fig:tree_interference}(a). Each beam splitter has two inputs and two outputs and is back-reflection-less. Electron keeps the direction of motion (transmits) with amplitude $t = \sqrt{T}$ and turns (reflects) with amplitude $r = i\sqrt{1-T}$. The particle can reach a given position by different paths, e.g., in order to reach link~2 at $\tau = 3$ it can move through link~1 at $\tau = 1$ and link~2 at $\tau=2$ (solid green arrows) or, alternatively, through link~0 at $\tau = 1$ and link~1 at $\tau = 2$ (dotted green arrows). The state of electron subsystem is defined by the $2\tau$ length vector at $\tau$\textsuperscript{th} step. The links can be enumerated by index $\zeta$ ($= \tau,j$), where~$j$ goes from~0 to $2\tau-1$. {In contrast to the previous model, the electron path for a given index $\zeta$ is indistinguishable.}
As before we place the spin at each link. It can rotate by some certain angle whenever the electron moves through correspondent link. Thereby, spin subsystem stores information about the electron path (which in principle can be measured later).
Two electron paths marked by rotated spins are depicted in Fig.~\ref{fig:tree_interference}(a) in solid and dotted red arrows. In this case the system can be described by~$2^\tau$ possible states as in the case without interference.

Present model is equivalent to the system with a one-dimensional electron and spin sets shown in Fig.~\ref{fig:tree_interference}(b). The latter recalls the very first model in Fig.~\ref{fig:tree}(a), however the spin set is replaced by the new one (in the initial state) after each step. In this setup the electron can interact with each spin only once and can interfere with itself.
Figure~\ref{fig:tree_interference}(c) shows the entropy as a function of time for beam splitter transmission probability $T=0.5$ and different~$\beta$'s. The same function for different transmission probabilities $T$ and $\beta=0.9$ is shown in Fig.~\ref{fig:tree_interference}(d).
The entropy is strictly growing in time. Introducing the interference in the system, which can interact with reservoir part only once does not break validity of H-theorem statement.
Note, when moving electron rotates spin by $\pi$ (full phase breaking case) the electron subsystem can be described by the master equation for probabilities and one can prove H-theorem analytically in the spirit of Boltzmann.

{\it Conclusion.} 
In this article we discussed several models of the random walk of the particle entangled with auxiliary system of spins. In the models with full phase breaking coming from the spin subsystem one can derive master equation for probabilities and, thus, prove the H-theorem explicitly. However, more complicated models with partial phase breaking can not be described by master equation for probabilities. The H-theorem statement still valid for some of them and becomes invalid for the others.

We have described three generic model types demonstrating both the monotonic entropy growth and non-monotonic entropy behavior. This behavior depends on the presence of the interference in the system and possibility of the multiple interactions with the reservoir parts.
The entropy in the system with intrinsic interference and multiple interactions with the reservoir parts increases in average, but shows non-monotonic behavior at small time scales. In systems interacting with reservoir parts only once the entropy monotonically increases and the H-theorem statement remains valid.
In our analysis we have considered quite general interaction properties omitting its details. Therefore, we believe the main assertions are valid for even wider class of models.

The work was supported by the RFBR Grant No. 11-02-00744-a (G.B.L.) and the U.S. Department of Energy Office of Science under the Contract No. DE-AC02-06CH11357 (I.A.S.).

\end{document}